\documentclass[12pt]{iopart}

\usepackage{amsfonts,amssymb,graphics,graphicx,bbm}
\usepackage[colorlinks=true,citecolor=blue,linkcolor=blue,urlcolor=blue]{hyperref}
\usepackage[usenames]{color}
\usepackage{graphicx}
\usepackage{subfigure}
\usepackage{bm}
\usepackage{color}
\usepackage{epstopdf}
\usepackage{amssymb}
\usepackage{amstext}
\usepackage{latexsym}
\usepackage{hyperref}
\usepackage{amsfonts}
\usepackage{psfrag}
\usepackage{graphicx}

\expandafter\let\csname equation*\endcsname\relax
\expandafter\let\csname endequation*\endcsname\relax
\usepackage{amsmath}

\newcommand{\ket}[1]{\vert #1 \rangle}
\newcommand{\bra}[1]{\langle #1 \vert}
\newcommand{\ketbra}[2]{\vert #1 \rangle \langle #2 \vert}
\newcommand{\braket}[2]{\langle #1 \vert #2 \rangle}

\newcommand{\abs}[1]{| #1 |}
\newcommand{\eg}{{\it{e.g.}}}

\begin{document}
\title[Quantum Reinforcement Learning for Eigensolver]{Reinforcement learning for semi-autonomous approximate quantum eigensolver}
\author{F. Albarr\'an-Arriagada $^{1,2,3}$, J. C. Retamal $^{2,3}$, E. Solano $^{1,4,5}$ and L. Lamata $^{4,6}$}
\address{$^1$ International Center in Quantum Artificial Intelligence for Science and Technology (QuArtist)  and Physics Department, Shanghai University, 200444 Shanghai, China}
\address{$^2$ Departamento de F\'isica, Universidad de Santiago de Chile (USACH), 
Avenida Ecuador 3493, 9170124, Santiago, Chile}
\address{$^3$ Center for the Development of Nanoscience and Nanotechnology 9170124, Estaci\'on Central, Santiago, Chile}
\address{$^4$ Department of Physical Chemistry, University of the Basque Country UPV/EHU, Apartado 644, 48080 Bilbao, Spain}
\address{$^5$ IKERBASQUE, Basque Foundation for Science, Maria Diaz de Haro 3, 48013 Bilbao, Spain}
\address{$^6$ Departamento de F\'isica At\'omica, Molecular y Nuclear, Universidad de Sevilla, 41080 Sevilla, Spain}
\ead{pancho.albarran@gmail.com}
\begin{abstract}
The characterization of an operator by its eigenvectors and eigenvalues allows us to know its action over any quantum state. Here, we propose a protocol to obtain an approximation of the eigenvectors of an arbitrary Hermitian quantum operator. This protocol is based on measurement and feedback processes, which characterize a reinforcement learning protocol. Our proposal is composed of two systems, a black box named environment and a quantum state named agent. The role of the environment is to change any quantum state by a unitary matrix $\hat{U}_E=e^{-i\tau\hat{\mathcal{O}}_E}$ where $\hat{\mathcal{O}}_E$ is a Hermitian operator, and $\tau$ is a real parameter. The agent is a quantum state which adapts to some eigenvector of $\hat{\mathcal{O}}_E$ by repeated interactions with the environment, feedback process, and semi-random rotations. With this proposal, we can obtain an approximation of the eigenvectors of a random qubit operator with average fidelity over 90\% in less than 10 iterations, and surpass 98\% in less than 300 iterations. Moreover, for the two-qubit cases, the four eigenvectors are obtained with fidelities above 89\% in 8000 iterations for a random operator, and fidelities of $99\%$ for an operator with the Bell states as eigenvectors. This protocol can be useful to implement semi-autonomous quantum devices which should be capable of extracting information and deciding with minimal resources and without human intervention.
\end{abstract}
\maketitle

\section{Introduction}
In the past few years, the symbiosis between quantum mechanics and machine learning into the topic named quantum machine learning (QML) has been a fruitful area \cite{Adcock2015,Biamonte2017Review,Dunjko2016,Dunjko2018Review}, either applying classical machine learning techniques to quantum tasks such as quantum metrology \cite{Hentschel2010,Hentschel2011}, quantum state estimation \cite{Torlai2018,Rocchetto2019}, and others \cite{Hase2017,Gupta2018,Gao2018,Bukov2018prb,Bukov2018prx,Melnikov2018}; or using quantum mechanics to enhance machine learning algorithms for classical  applications \cite{Aimeur2013,Lloyd2013,Rebentrost2014,Li2015,Cai2015,Dunjko2016,Sheng2017,Schuld2019}. Any machine learning algorithm can be classified into learning from big data and learning from interactions. 

For the first group, we have two classes of algorithms, one of them are the supervised learning algorithms, which use a previously labeled data set named training data to infer a labeled criterion which is used to classify new data; a remarkable example is pattern recognition algorithms \cite{Jain2007, Carrasquilla2017, Schutzhold2003}. The other class is unsupervised learning algorithms. In this case, the training data is not necessary, and the approach is to group the unlabeled data in different sets, where each set is characterized by the mean value of some property of its constituents. The different groups are constructed to optimize some indicator of the dispersion in each subset with respect to the value that characterized it, \eg, the standard deviation. An example of these algorithms is the clustering problem \cite{Fahad2014, Otterbach2017}. 

For the second group, we have the reinforcement learning (RL) algorithms \cite{Sutton1998Book}. Here, one accessible and manipulable system called agent $(A)$ interacts with another unknown system called environment $(E)$. The strategy relies on $A$ improving its performance in a specific task $\mathcal{Q}(A, E)$, which depends on the state of the systems $A$ and $E$. This improvement employs the results of multiple interactions among $A$ and $E$. The general framework of the RL paradigm is composed of three parts, the policy, the reward function (RF) and the value function (VF). The policy defines the main steps of the algorithm that we can divide into three steps. First, the information extraction, which considers the interaction among $A$ and $E$, and how to obtain the information from it. Second, the feedback loop, that specifies the channel used to communicate the information extracted to $A$. Third, the decision process, where we decide the action on $A$ in order to progress towards the aimed-for goal, and then start with the information extraction again. The RF defines the criterion to reward (punish) the actions which improve (worsen) the performance of $A$ respect to the task $\mathcal{Q}(A, E)$ at each step. Finally, the VF gives us the global performance of the algorithm, ensuring the convergence of it. One of the most impressive examples of this paradigm is the recent developing of chess, go and shogi masters players without database \cite{Silver2017, Silver2018}. This class of algorithms mimic the most primitive form of human learning, commonly named trial and error. It means that a near-future implementation of quantum artificial intelligence may apply this paradigm to a quantum system to enhance a quantum task as the main way to learn. For this reason, the development of the quantum version of the RL paradigm has played an important role in QML in recent years \cite{DongChen,Paparo2014,Dunjko2016,Lamata2017,CardenasLopez2018,Crawford2019}.

A crucial task in physics is the characterization of the different interactions among systems. This characterization is helpful to evaluate the risks of our actions and act to minimize them. Therefore, any autonomous artificial intelligence must have this ability. 

In quantum mechanics, a physical interaction (observable) is represented by a Hermitian matrix or quantum operator, which is characterized by its eigenvalues and eigenvectors. The calculation of the eigenvectors and eigenvalues of a quantum interaction by a classical computer implies that we need to encode the quantum information into classical bits, which is inconvenient for unknown quantum interactions. Moreover, the implementation of a full quantum eigensolver \cite{Abrams1999, Jaksch2003, Wang2010, Wang2016} using near-future quantum computers seems impractical due to the number of needed resources \cite{Wecker2014}. The emergence of hybrid classical-quantum algorithms in the past few years \cite{Peruzzo2014, YungLamata2014,McClean2016, OMalley2016, Kandala2017, Hempel2018, Kokail2019} opens the door to the development of useful eigensolvers. Nevertheless, these works are mainly focused on the eigenvalues, eigenvectors, and properties of quantum systems such as molecules, being the characterization of a physical interaction less studied. 

In this article, we propose a hybrid quantum-classical algorithm to calculate an approximation to the eigenvector of any quantum interaction described by a Hermitian matrix with minimal resources \cite{Codes}. In our proposal, we use single-shot measurement and classical communication given by a feedback loop, which characterizes a RL protocol. The main goal of this proposal is to obtain a high-fidelity approximation (above 98\% for the single-qubit case), without measuring fidelities or some expectation value, which reduce drastically the number of iterations of the algorithm, decreasing the effect of noise sources, and without human intervention. We also show how to extend the algorithm to the multiqubit and high-dimensional situations.  This protocol could be useful to implement semi-autonomous quantum devices with the capability to decide using the characterization of an interaction, which is an essential ingredient for the implementation of artificial quantum intelligence \cite{Dunjko2018Review} and artificial quantum life \cite{AlvarezRodriguez2016, AlvarezRodriguez2018}.


\section{Quantum eigensolver protocol}
Our proposal is related to recent works about a measurement-based algorithm to adapt one known state to another unknown one \cite{AlbarranArriagada2018, Yu2019,Olivares2018}. Here, we define the general framework of our protocol based on the RL paradigm and then, we explain in details the single qubit case, the single qudit case, and the multiqubit case.

In our protocol, we consider as the agent a manipulable and known quantum system described by the state $\ket{\phi_{A,0}}$, which correspond to any initialization of a given physical system. The environment is a black box, which produces an unknown interaction inside it. This interaction is characterized by an unknown Hermitian operator $\hat{\mathcal{O}}_E$, which generates a unitary transformation $\hat{U}_E=e^{-i\tau\hat{\mathcal{O}}_E}$ over the quantum system $A$ when it interacts with the system $E$, where $\tau$ is a parameter related to the interaction time with the black-box, \eg, a spin particle (agent) traversing a region with a magnetic field (environment) for a time $t\sim\tau$.

\begin{figure}[t]
	\centering
	\includegraphics[width=0.6\linewidth]{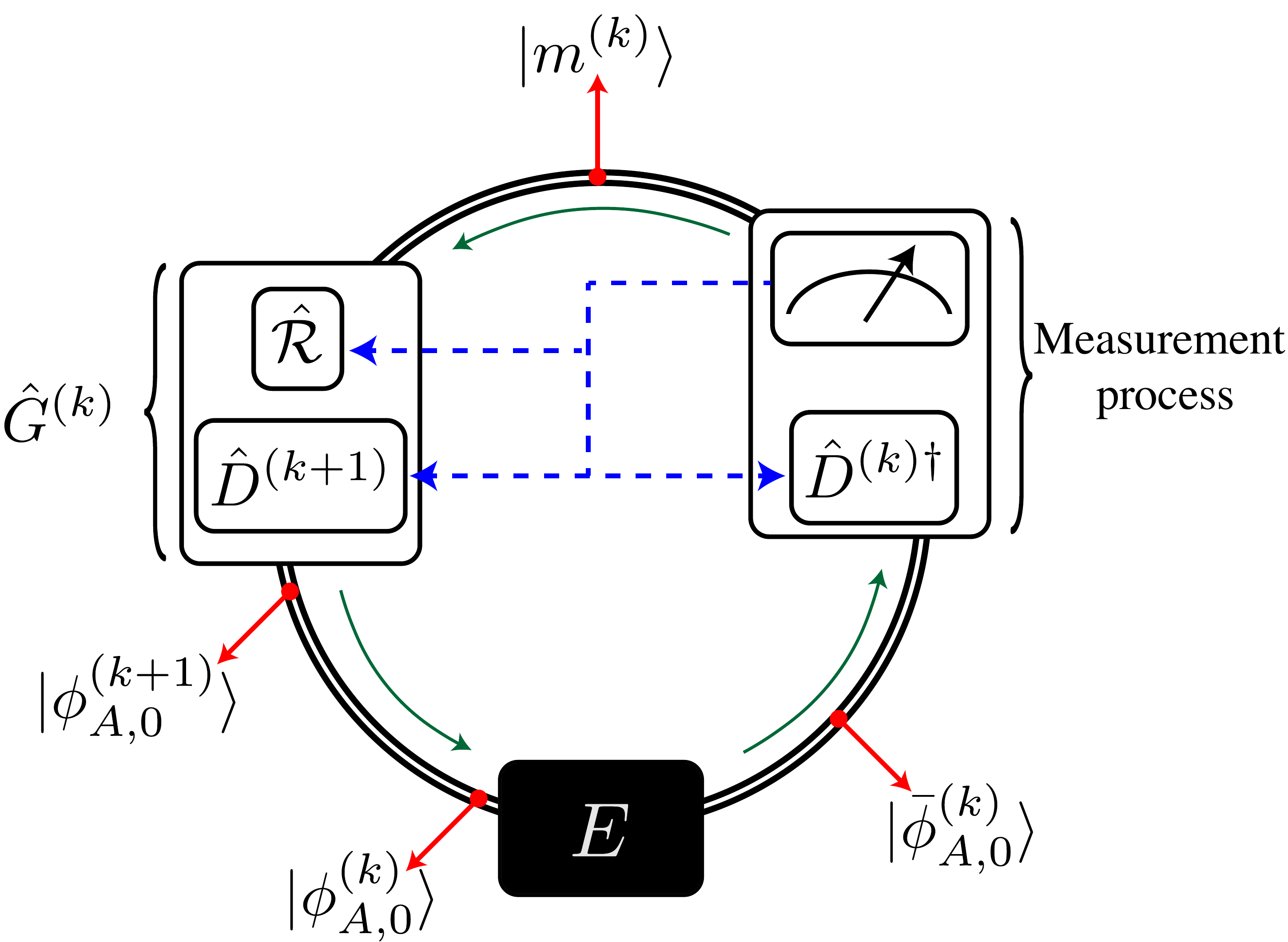}
	\caption{Diagram of the protocol. The solid green arrows show flow direction of $A$ state. The blue dashed arrows represent the feedback loops, and the red arrow with dot end marks the states in each step. The state $\ket{\phi_{A,0}^{(k)}}$ corresponds to the start point of the $k$th iteration, and the state $\ket{\phi_{A,0}^{(k+1)}}$ corresponds to the end point of the $k$th iteration, that is also the state at the beginning of the $(k+1)$th iteration.}
	\label{Fig1}
\end{figure}

The policy is as follows: 
\begin{itemize}
    \item \textit{Information extraction}: The system $A$ interacts with $E$ changing its state as 
    \begin{equation}
        \ket{\bar{\phi}_{A,0}}=\hat{U}_E\ket{\phi_{A,0}}.
        \label{Eq01}
    \end{equation}
	Next, we perform a measurement process over $\ket{\bar{\phi}_{A,0}}$ in the basis $\{\ket{\phi_{A,0}},...,\ket{\phi_{A,d-1}}\}$, where $d$ is the dimension of the Hilbert space of $A$ and $\braket{\phi_{A,j}}{\phi_{A,k}}=\delta_{j,k}$.
    
    \item \textit{Feedback loop}: The information of the measuring process is communicated to a command center with the ability to perform a unitary transformation $\hat{\mathcal{U}}_j$ (quantum gate) over the state of $A$ in order to change the possible results in the next information extraction step.
    
    \item \textit{Decision process}: If the outcome of the measurement process is the state $\ket{\phi_{A,j}}$, with $j\ne 0$, this means that $\ket{\phi_{A,0}}$ changes when system $A$ interacts with $E$, therefore, $\ket{\phi_{A,0}}$ cannot be an eigenvector of $\hat{\mathcal{O}}_E$. In this case, we define the unitary transformation $\hat{\mathcal{U}}_j$ as
\begin{equation}
    \hat{\mathcal{U}}_j=e^{-i\varphi_y\hat{S}_{y,j}}e^{-i\varphi_z\hat{S}_{z,j}}e^{-i\varphi_x\hat{S}_{x,j}},
    \label{Eq02}
\end{equation}
where
\begin{eqnarray}
    \hat{S}_{x,j}&=&\frac{1}{2}\left(\ketbra{\phi_{A,0}}{\phi_{A,j}}+\ketbra{\phi_{A,j}}{\phi_{A,0}}\right),\nonumber\\
    \hat{S}_{y,j}&=&-\frac{i}{2}\left(\ketbra{\phi_{A,0}}{\phi_{A,j}}-\ketbra{\phi_{A,j}}{\phi_{A,0}}\right),\nonumber\\
    \hat{S}_{z,j}&=&\frac{1}{2}\left(\ketbra{\phi_{A,0}}{\phi_{A,0}}-\ketbra{\phi_{A,j}}{\phi_{A,j}}\right),
    \label{Eq03}
\end{eqnarray}
and $\varphi_{\alpha}$ is a random angle in the range $[-w\pi,w\pi]$, with $w$ the searching range given by the RF. We note that $\hat{\mathcal{U}}_j$ is a pseudo random rotation in the subspace expanded by $\{\ket{\phi_{A,0}},\ket{\phi_{A,j}}\}$. For this outcome we define the state of $A$ as $\hat{\mathcal{U}}_j\ket{\phi_{A,0}}$, and start again with the information extraction step.
    
    If the outcome of the measuring process is $\ket{\phi_{A,0}}$, it means that $\ket{\phi_{A,0}}$ could be an eigenvector of $\hat{\mathcal{O}}_E$. We point out that the eigenvectors of an operator remain constant up to a global phase under the action of a function of this operator. In this case, we apply the identity operator $\mathbb{I}$. Moreover, we keep the same state $\ket{\phi_{A,0}}$ and start again with the information extraction step. Figure~\ref{Fig1} shows a scheme of the policy of the algorithm.
\end{itemize}

For the RF we define the \textit{reward rate} $r<1$ and the \textit{punishment rate} $p>1$. If the outcome of the measure is $\ket{\phi_{A,0}}$ we define $\bar{w}=w\cdot r$ and  $\bar{w}=w\cdot p$ in other case. Finally, we renamed $w=\bar{w}$ for the next iteration of the algorithm, which means that when we measure $\ket{\phi_{A,0}}$ we reduce the searching range, and we increase it in other case. The initial value for $w$ is chosen according to the problem.

As we can note, the protocol does not need store the states, or all the history of the algorithm, it only needs to store the final operation $\hat{D}^{(N)}$ via storing the parameters that characterize this operation classically.

To ensure the convergence of our algorithm, we define the VF as the value of $w$. This implies that, when $w\rightarrow 0$, our protocol converges. For a correct choice of $r$ and $p$ we have that $w\rightarrow 0$ only if we obtain, in the measurement process of $\ket{\bar{\phi}_{A,0}}$, the outcome $\ket{\phi_{A,0}}$ many times in a row. This means that $\braket{\phi_{A,0}}{\bar{\phi}_{A,0}}\sim 1$, therefore $\ket{\phi_{A,0}}$ is an approximate eigenvector of $\hat{\mathcal{O}}_E$.

As this is an iterative protocol, we define the following notation for the remainder of the article: any super-index between parenthesis refers to the iteration of the algorithm, \eg, $\ket{\phi_{A,0}^{(4)}}$ is the state of $A$ before the interaction with $E$ in the fourth iteration. Similarly, $\hat{\mathcal{U}}^{(k)}_j$ is the unitary transformation defined in the decision process for the iteration $k$. As a special case, the super-index $(1)$ refers to the initial values, \eg, $w^{(1)}$ represents the initial searching range.

It is necessary to mention that our algorithm uses one single-shot measurement per loop, representing advantage with respect to employing an expectation value or the fidelity. The latter imply hundreds of measurements for a two-level system, being this proposal exposed less time to noise sources. Also, as we use pseudo-random operations $\hat{D}^{(k)}$, the effect of any noise in the gate can be seen as part of the randomness of the protocol.


\subsection{Single-qubit case}
In the single-qubit case, $\hat{\mathcal{O}}_E$ is described by a $2\times2$ Hermitian matrix with eigenvectors $\{\ket{v_0},\ket{v_1}\}$ and eigenvalues $\{\lambda_0,\lambda_1\}$ respectively. As these two eigenvectors are orthonormal, we can write
\begin{eqnarray}
	\ket{v_0}&=&\cos\left(\frac{\alpha}{2}\right)\ket{0}+e^{i\beta}\sin\left(\frac{\alpha}{2}\right)\ket{1},\nonumber\\
	\ket{v_1}&=&\sin\left(\frac{\alpha}{2}\right)\ket{0}-e^{i\beta}\cos\left(\frac{\alpha}{2}\right)\ket{1}
	\label{Eq04}
\end{eqnarray}
where $\alpha\in[0,2\pi]$, $\beta\in[0,\pi]$ and
\begin{equation}
	\ket{0}=\begin{pmatrix}
		1\\
		0
	\end{pmatrix},\quad\ket{1}=\begin{pmatrix}
		0\\
		1
	\end{pmatrix}.
	\label{Eq05}
\end{equation}

We define $\hat{\mathcal{O}}_E$ and $\hat{U}_E$ as
\begin{eqnarray}
	\hat{\mathcal{O}}_E&=&\lambda_0\ketbra{v_0}{v_0}+\lambda_1\ketbra{v_1}{v_1},\nonumber\\
	\hat{U}_E&=&e^{-i\lambda_0\tau}\ketbra{v_0}{v_0}+e^{-i\lambda_1\tau}\ketbra{v_1}{v_1}.
	\label{Eq06}
\end{eqnarray}

\textit{Policy}. In this case, we write the state $\ket{\phi_{A,0}^{(k)}}$ before the black-box as
\begin{equation}
	\ket{\phi_{A,0}^{(k)}}=\cos\left(\frac{\theta^{(k)}}{2}\right)\ket{0}+e^{i\varphi^{(k)}}\sin\left(\frac{\theta^{(k)}}{2}\right)\ket{1},
	\label{Eq07}
\end{equation}
and the state $\ket{\bar{\phi}_{A,0}^{(k)}}$ after $E$ as
\begin{eqnarray}
	&&\ket{\bar{\phi}_{A,0}^{(k)}}=\cos\left(\frac{\bar{\theta}^{(k)}}{2}\right)\ket{0}+e^{i\bar{\varphi}^{(k)}}\sin\left(\frac{\bar{\theta}^{(k)}}{2}\right)\ket{1}\nonumber\\
	&&=\cos\left(\frac{\Delta^{(k)}_{\theta}}{2}\right)\ket{\phi_{A,0}^{(k)}}+e^{i\Delta_{\varphi}^{(k)}}\sin\left(\frac{\Delta_{\theta}^{(k)}}{2}\right)\ket{\phi_{A,1}^{(k)}}
	\label{Eq08}
\end{eqnarray}
where
\begin{eqnarray}
	&&\ket{\phi_{A,1}^{(k)}}=\sin\left(\frac{\theta^{(k)}}{2}\right)\ket{0}-e^{i\varphi^{(k)}}\cos\left(\frac{\theta^{(k)}}{2}\right)\ket{1}.
	\label{Eq09}
\end{eqnarray}
For the explicit form $\bar{\theta}^{(k)}$ and $\bar{\phi}^{(k)}$ in terms of $\alpha$, $\beta$, $\tau$ and the eigenvalues of $\hat{\mathcal{O}}_E$ see appendix \ref{appA}. Moreover, for the explicit form of $\Delta_{\theta}^{(k)}$ and $\Delta_{\phi}^{(k)}$, see appendix \ref{appB}. Now, to perform the measurement process over $\ket{\bar{\phi}_{A,0}^{(k)}}$, we apply the basis-rotation matrix
\begin{equation}
	\hat{D}^{(k)\dagger}=\ketbra{0}{\phi_{A,0}^{(k)}}+\ketbra{1}{\phi_{A,1}^{(k)}},
	\label{Eq10}
\end{equation}
in order to measure in the basis $\{\ket{0}, \ket{1}\}$ for all iterations. After the measurement process, the state of $A$ is $\ket{m^{(k)}}$, where $m^{(k)}\in\{0,1\}$ is the outcome of the measurement with probabilities $\mathcal{P}_0^{(k)}=\cos^2(\Delta^{(k)}/2)$ and $\mathcal{P}_1^{(k)}=\sin^2(\Delta^{(k)}/2)$, respectively. If $m^{(k)}=0$, then we transform the state $\ket{0}\rightarrow\ket{\phi_{A,0}^{(k)}}$, using the matrix $\hat{D}^{(k)}$, and start again the algorithm. If $m^{(k)}=1$, we  transform the state $\ket{1}\rightarrow\ket{\phi_{A,0}^{(k)}}$ using $\hat{D}^{(k)}\sigma_x$, where $\sigma_x$ is the Pauli matrix $x$, and apply the pseudo-random operator  $\hat{\mathcal{U}}_1^{(k)}$ defined by Eq. (\ref{Eq02}). Then, after the measurement process, we apply over $\ket{m^{(k)}}$ the operator $\hat{G}^{(k)}_0$ defined by
\begin{equation}
	\hat{G}^{(k)}_0=\hat{D}^{(k+1)}\hat{\mathcal{R}}
	\label{Eq11}
\end{equation}
where
\begin{eqnarray}
	\hat{D}^{(k+1)}&=&(1-m^{(k)})\hat{D}^{(k)}+m^{(k)}\hat{\mathcal{U}}_1^{(k)}\hat{D}^{(k)},\nonumber\\
	\hat{\mathcal{R}}&=&(1-m^{(k)})\mathbb{I}+m^{(k)}\sigma_x.
\label{Eq12}
\end{eqnarray}
Given that $\hat{D}^{(k)}$ transforms $\ket{\phi_{A,j}^{(k)}}\rightarrow\ket{j}$ ($\ket{j}\in\{\ket{0},\ket{1}\}$), we can write $\hat{\mathcal{U}}_1^{(k)}=\hat{D}^{(k)}\hat{u}_1\hat{D}^{(k)\dagger}$, where
\begin{equation}
	\hat{u}_1=e^{-i\varphi_y\hat{S}_{y}}e^{-i\varphi_z\hat{S}_{z}}e^{-i\varphi_x\hat{S}_{x}},
	\label{Eq13}
\end{equation}
with $\hat{S}_{j}=(1/2)\sigma_j$ the spin operators, with $\sigma_j$ the Pauli matrix $j$. Then, the operator $\hat{D}^{(k+1)}$ reads
\begin{equation}
	\hat{D}^{(k+1)}=(1-m^{(k)})\hat{D}^{(k)}+m^{(k)}\hat{D}^{(k)}\hat{u}_1.
\label{Eq14}
\end{equation}

For this case, the RF that defines the value of $w^{(k+1)}$ for each step reads
\begin{equation}
	w^{(k+1)}=\left[(1-m^{(k)})r + m^{(k)}p\right] w^{(k)},
	\label{Eq15}
\end{equation}
where $r$ and $p$ are the reward rate and punishment rate, respectively described previously.

When the algorithm converges, we have $\ket{\phi_{A,0}^{(N)}}\approx\ket{\bar{\phi}_{A,0}^{(N)}}$, where $N$ is the number of iterations. Moreover, in this case $\hat{D}^{(N)}$ is an approximation of the matrix that diagonalizes $\hat{\mathcal{O}}_E$, that is
\begin{equation}
	\hat{D}^{(N)\dagger}\hat{\mathcal{O}}_E\hat{D}^{(N)}\sim\lambda_0\ketbra{0}{0} + \lambda_1\ketbra{1}{1}.
	\label{Eq16}
\end{equation}

In order to explore the complete space we must choose $w^{(1)}=1$.


\subsection{Single-qudit case}
In this case, the agent is a $d$-dimensional system or qudit, the operator $\hat{\mathcal{O}}_E$ is described by a $d\times d$ Hermitian matrix with eigenvalues $\{\lambda_j\}$, eigenvectors $\{\ket{v_j}\}$ and $j=\{0,1,2,...,d-1\}$. In the $k$th iteration of the algorithm, the state of $A$ before $E$ reads
\begin{equation}
	\ket{\phi_{A,0}^{(k)}}=\sum_{j=0}^{d-1}c_j\ket{j},
	\label{Eq17}
\end{equation}
while for simplicity we choose the initial state $\ket{\phi_{A,0}^{(1)}}=\ket{0}$. After the interaction with $E$, we have
\begin{equation}
	\ket{\bar{\phi}_{A,0}^{(k)}}=\hat{U}_E\ket{\phi_{A,0}^{(k)}}=\sum_{j=0}^{d-1}\bar{c}_j\ket{\phi_{A,j}^{(k)}}.
	\label{Eq18}
\end{equation}

Subsequently, we apply the operator $\hat{D}^{(k)\dagger}$, which is defined now as
\begin{equation}
	\hat{D}^{(k)\dagger}=\sum_{j=0}^{d-1}\ketbra{j}{\phi_{A,j}^{(k)}},
	\label{Eq19}
\end{equation}
and perform the measurement process in the basis $\{\ket{0},\ket{1},...,\ket{d-1}\}$. After this process, the state of $A$ is $\ket{m^{(k)}}$, where $m^{(k)}\in\{0,1,...,d-1\}$ is the outcome of the measurement process. In this case the decision process applies the operator $\hat{G}^{(k)}_0$ defined by Eq. (\ref{Eq11}), but with
\begin{eqnarray}
	\hat{\mathcal{R}}=\delta_{0,m^{(k)}}(\mathbb{I}-\hat{\mathcal{X}})+\hat{\mathcal{X}},\nonumber\\
	\hat{D}^{(k+1)}=\sum_{j=0}^{d-1}\delta_{j,m^{(k)}}\hat{\mathcal{U}}_j^{(k)}\hat{D}^{(k)},
	\label{Eq20}
\end{eqnarray}
where 
\begin{eqnarray}
	\hat{\mathcal{X}}=\sum_{j=1}^{d-1}\left(\ketbra{0}{j}+\ketbra{j}{0}\right)
	\label{Eq21}
\end{eqnarray}
with $\hat{\mathcal{U}}_{m^{(k)}}^{(k)}$ as defined in Eq. (\ref{Eq02}) and $\hat{\mathcal{U}}_{0}^{(k)}=\mathbb{I}$. Also in this case $\hat{\mathcal{U}}_j^{(k)}=\hat{D}^{(k)}\hat{u}_j\hat{D}^{(k)\dagger}$, where
\begin{equation}
	\hat{u}_j=e^{-i\varphi_y\hat{S}_{y}^j}e^{-i\varphi_z\hat{S}_{z}^j}e^{-i\varphi_x\hat{S}_{x}^j},
	\label{Eq22}
\end{equation}
and 
\begin{eqnarray}
	\hat{S}_{x}^j&=&\frac{1}{2}\left(\ketbra{0}{j}+\ketbra{j}{0}\right),\nonumber\\
	\hat{S}_{y}^j&=&-\frac{i}{2}\left(\ketbra{0}{j}-\ketbra{j}{0}\right),\nonumber\\
	\hat{S}_{z}^j&=&\frac{1}{2}\left(\ketbra{0}{0}+\ketbra{j}{j}\right),
	\label{Eq23}
\end{eqnarray}
therefore,
\begin{equation}
	\hat{D}^{(k+1)}=\sum_{j=0}^{d-1}\delta_{j,m^{(k)}}\hat{D}^{(k)}\hat{u}_{m^{(k)}}.
	\label{Eq24}
\end{equation}

The state of $A$ for the next iteration reads $\ket{\phi_{A,0}^{(k+1)}}=\hat{G}^{(k)}_0\ket{m^{(k)}}$. 

Finally, the RF that updates the value of the searching range is given by
\begin{equation}
	w^{(k+1)}=\left[(r-p)\delta_{0,m^{(k)}}+p\right]w^{(k)}.
	\label{Eq25}
\end{equation}

Once the algorithm converges, we have that 
\begin{equation}
	\ket{\phi_{A,0}^{(N_0+1)}}=\hat{D}^{(N_0)}\ket{\phi_{A,0}^{(1)}},
	\label{Eq26}
\end{equation}
is an approximate eigenvector, therefore,
\begin{equation}
	\abs{\bra{\phi_{A,0}^{(N_0+1)}}\hat{\mathcal{O}}_E\ket{\phi_{A,0}^{(N_0+1)}}}\sim 1.
	\label{Eq27}
\end{equation}

In order to find another eigenvector of $\hat{\mathcal{O}}_E$, we start again the algorithm for the iteration $N_0+1$, i.e., $w^{(N_0+1)}=w^{(1)}=2\pi$, but now the state before $E$ is given by $\ket{\phi_{A,1}^{(N_0+1)}}=\hat{D}^{(N_0)}\ket{\phi_{A,1}^{(1)}}$. We redefine Eq. (\ref{Eq23}) as 
\begin{eqnarray}
	\hat{S}_{x}^j&=&\frac{1}{2}\left(\ketbra{1}{j}+\ketbra{j}{1}\right),\nonumber\\
	\hat{S}_{y}^j&=&-\frac{i}{2}\left(\ketbra{1}{j}-\ketbra{j}{1}\right),\nonumber\\
	\hat{S}_{z}^j&=&\frac{1}{2}\left(\ketbra{1}{1}-\ketbra{j}{j}\right).
	\label{Eq28}
\end{eqnarray}
Thus, we can calculate the operator $\hat{u}_j$ as in Eq. (\ref{Eq22}). 

The decision process changes as
\begin{equation}
	\hat{G}^{(k)}_1=\hat{D}^{(k+1)}\hat{\mathcal{R}}_1,
	\label{Eq29}
\end{equation}
where
\begin{eqnarray}
	\hat{\mathcal{R}}_1=\delta_{1,m^{(k)}}(\mathbb{I}-\hat{\mathcal{X}}_1)+\hat{\mathcal{X}}_1,\nonumber\\
	\hat{D}^{(k+1)}=\sum_{j=0}^{d-1}\delta_{j,m^{(k)}}\hat{D}^{(k)}\hat{u}_j,\nonumber\\
	\hat{\mathcal{X}}_1=\sum_{j\ne1}\left(\ketbra{1}{j}+\ketbra{j}{1}\right),
	\label{Eq30}
\end{eqnarray}
and $\hat{u}_0=\hat{u}_1=\mathbb{I}$. Finally, the RF reads,
\begin{equation}
	w^{(k+1)}=\left[(r-p)\delta_{1,m^{(k)}}-p\delta_{0,m^{(k)}}+p\right]w^{(k)}.
	\label{Eq31}
\end{equation}

These changes mean that we perform the protocol in the subspace orthogonal to $\ket{\phi_{A,0}^{(1)}}$. When the algorithm converges again, after $N_1$ iterations more, we have that the states $\hat{D}^{(N_0+N_1)}\ket{\phi_{A,0}^{(1)}}$ and $\hat{D}^{(N_0+N_1)}\ket{\phi_{A,1}^{(1)}}$ are approximate eigenvectors. Therefore, to obtain the next eigenvector we perform the algorithm again but in the subspace orthogonal to $\{\ket{\phi_{A,0}^{(1)}},\ket{\phi_{A,1}^{(1)}}\}$, and so on. At $N=N_0+N_1+...+N_{d-2}$ iterations we have that the states $\ket{\phi_{A,j}^{N}}=\hat{D}^{(N)}\ket{\phi_{A,j}^{(1)}}$ with $j=0,1,...,d-1$ are the $d$ eigenvectors of $\hat{\mathcal{O}}_E$. 


\subsection{Multiqubit case}
For this case, we can suppose that the system $A$ is a qudit state, where now the states $\ket{j}$ of the basis, correspond to the binary representation of $j$ with $log_2(d)$ digits. For example, for $d=16$ we have $4$ digits, where each of them represents the state of a qubit; then $\ket{5}=\ket{0101}$. Also, we can produce the different operators $\hat{u}_j$ using controlled-not gates and single-qubit rotations \cite{Nielsen2010Book}. Therefore, we can map this problem to the qudit case obtaining the same algorithm as in the previous case.\\

As we can see from this section, our protocol does not need to encode quantum information in a classical processor, being advantageous with respect to classical algorithms that need to characterize the quantum interactions by quantum tomography. The latter imply hundreds of measurements of the quantum system, using in this process more resources than the entire algorithm proposed. Moreover, as our algorithm finds the eigenstate statistically, it is simpler than a full quantum algorithm that finds the eigenstates exactly, being our protocol experimentally feasible. The references \cite{Yu2019, Olivares2018} show the experimental implementation of an algorithm that employs the same basics steps in which our current algorithm is based, for the case of quantum states, instead of quantum operators, opening the door to the implementation of this work.

\begin{figure}[t]
	\centering
	\includegraphics[width=0.6\linewidth]{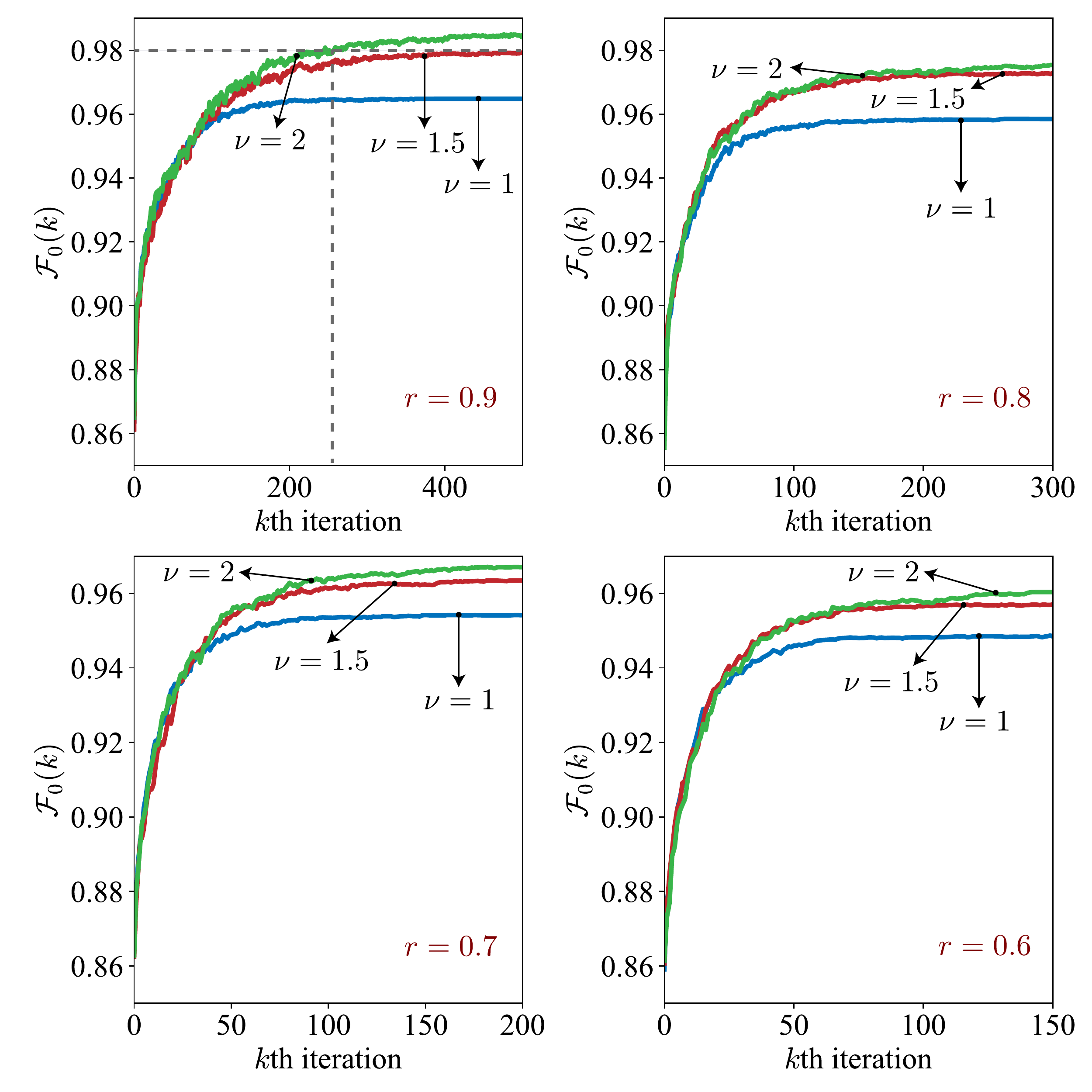}
	\caption{Numerical results for the mean fidelity $\mathcal{F}_0(k)$ given by Eq. (\ref{Eq33}) where $\hat{\mathcal{O}}_E$ corresponds to a random Hermitian matrix acting over a single qubit. We employ $\mathcal{N}=1000$.}
	\label{Fig3}
\end{figure}

\begin{figure}[b]
	\centering
	\includegraphics[width=0.6\linewidth]{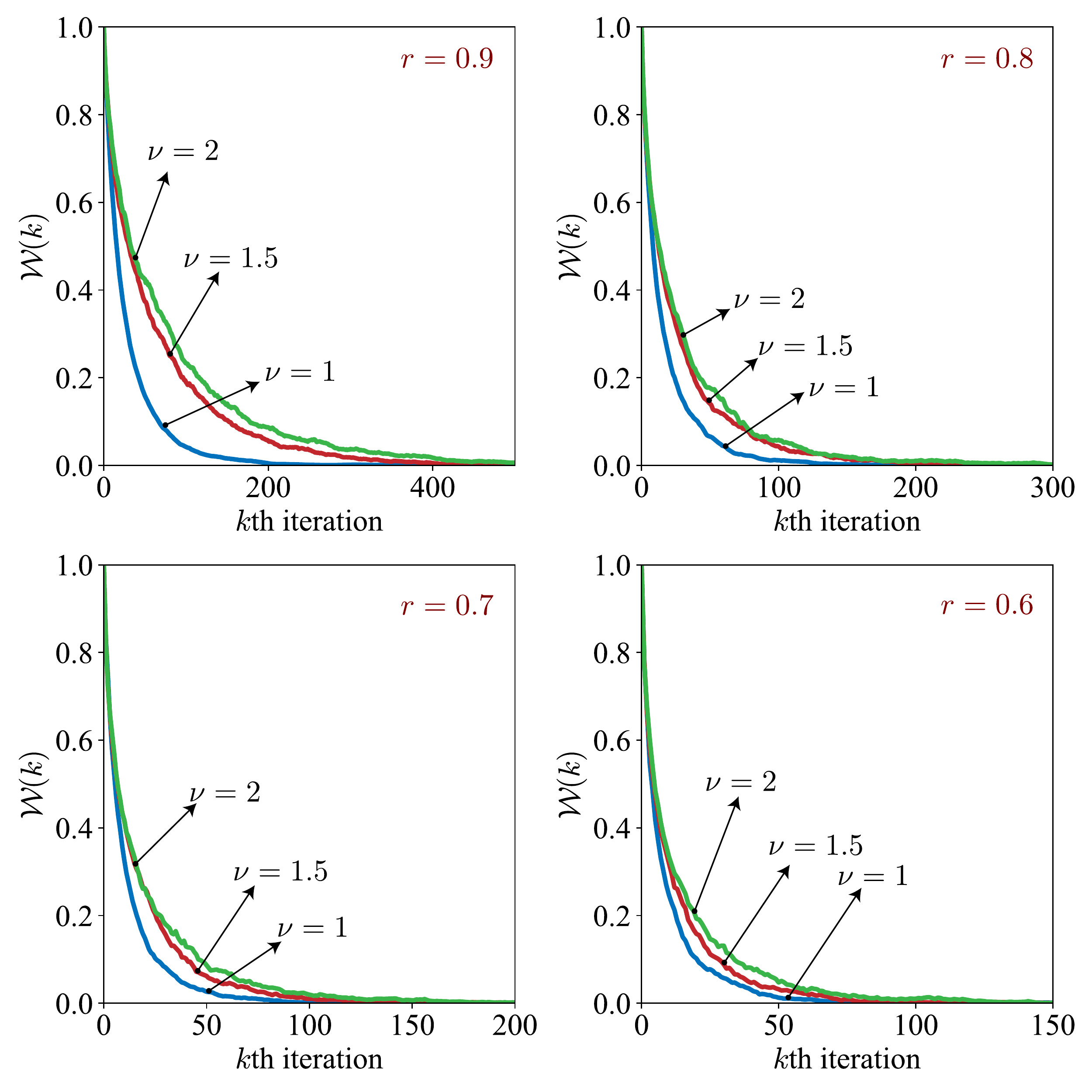}
	\caption{Numerical results for the mean searching rate $\mathcal{W}(k)$ given by Eq. (\ref{Eq33}) where $\hat{\mathcal{O}}_E$ corresponds to a random Hermitian matrix acting over a single-qubit. We employ $\mathcal{N}=1000$.}
	\label{Fig4}
\end{figure}
\section{Numerical results}
It is convenient to define the following quantities for the numerical analysis of the protocol, $\nu=r\cdot p \Rightarrow p=\nu/r$, with $r$ ($p$) the reward (punishment) rate, the total number of rewards  $n_r$ and the total number of punishments $n_p$ in the algorithm. The VF of our algorithm is the value of $w^{(N)}=r^{n_r}p^{n_p}$ where $N=n_r+n_p$ are the total number of iterations. Also, we can rewrite
\begin{equation}
w^{(N)}=r^{n_r-n_p}\nu^{n_p},
\label{Eq32}
\end{equation}
where the convergence condition is given by $w^{(N)}\ll1$. If $\nu<1$, we see from Eq. (\ref{Eq32}) that the convergence condition can be satisfied even if $n_p\sim n_r$, which implies that the protocol does not necessarily converge to the eigenstates of $\hat{\mathcal{O}}_E$. If $\nu=1$, we have that $w^{(N)}\rightarrow 0\iff n_r\gg n_p$. For $\nu>1$, the algorithm converges whenever $n_r\ggg n_p$. Moreover, when $\nu$ is larger, the algorithm needs more iterations to converge, but nevertheless it achieves larger fidelities. This is the exploration versus exploitation balance known in reinforcement learning. Here, we perform the simulation for a single- and two-qubit case for different values of $\nu$ and $r$. Remember that for all cases we choose $w^{(1)}=1$. Also, for simplicity we choose $\ket{\phi_{A,0}^{(1)}}=\ket{0}$ for the single-qubit case and $\ket{\phi_{A,j}^{(1)}}=\ket{j_{\textrm{bin}}}$ for the two-qubit case, where $j_{\textrm{bin}}$ is the binary representation of $j$, \eg, $\ket{\phi_{A,2}^{(1)}}=\ket{10}$. Moreover, $\hat{D}^{(1)}=\mathbb{I}$ for all cases.

Finally, as the unitary operator $\hat{u}_j$ given by Eq. (\ref{Eq22}) depends on pseudo-randoms angles, we perform many times the algorithm, defining the mean fidelity $\mathcal{F}$ and the mean searching range $\mathcal{W}$ as

\begin{eqnarray}
	\mathcal{F}_j(k)&=&\max_{\ell}\frac{1}{\mathcal{N}}\sum_{i=1}^{\mathcal{N}}|\bra{\ell_E}\hat{D}_i^{(k)}\ket{j}|,\nonumber\\
	\mathcal{W}(k)&=&\frac{1}{\mathcal{N}}\sum_{i=1}^{\mathcal{N}}w^{(k)}_i,
	\label{Eq33}
\end{eqnarray}
where $\ket{\ell_E}$ is the $\ell$th eigenvector of $\hat{\mathcal{O}}_E$, the index $i$ refers to the $i$th repetition of the protocol and $\mathcal{N}$ is the total number of repetitions. In all subsequent cases we choose $\mathcal{N}=1000$.

\subsection{Single-qubit case}
\begin{figure}[t]
	\centering
	\includegraphics[width=0.6\linewidth]{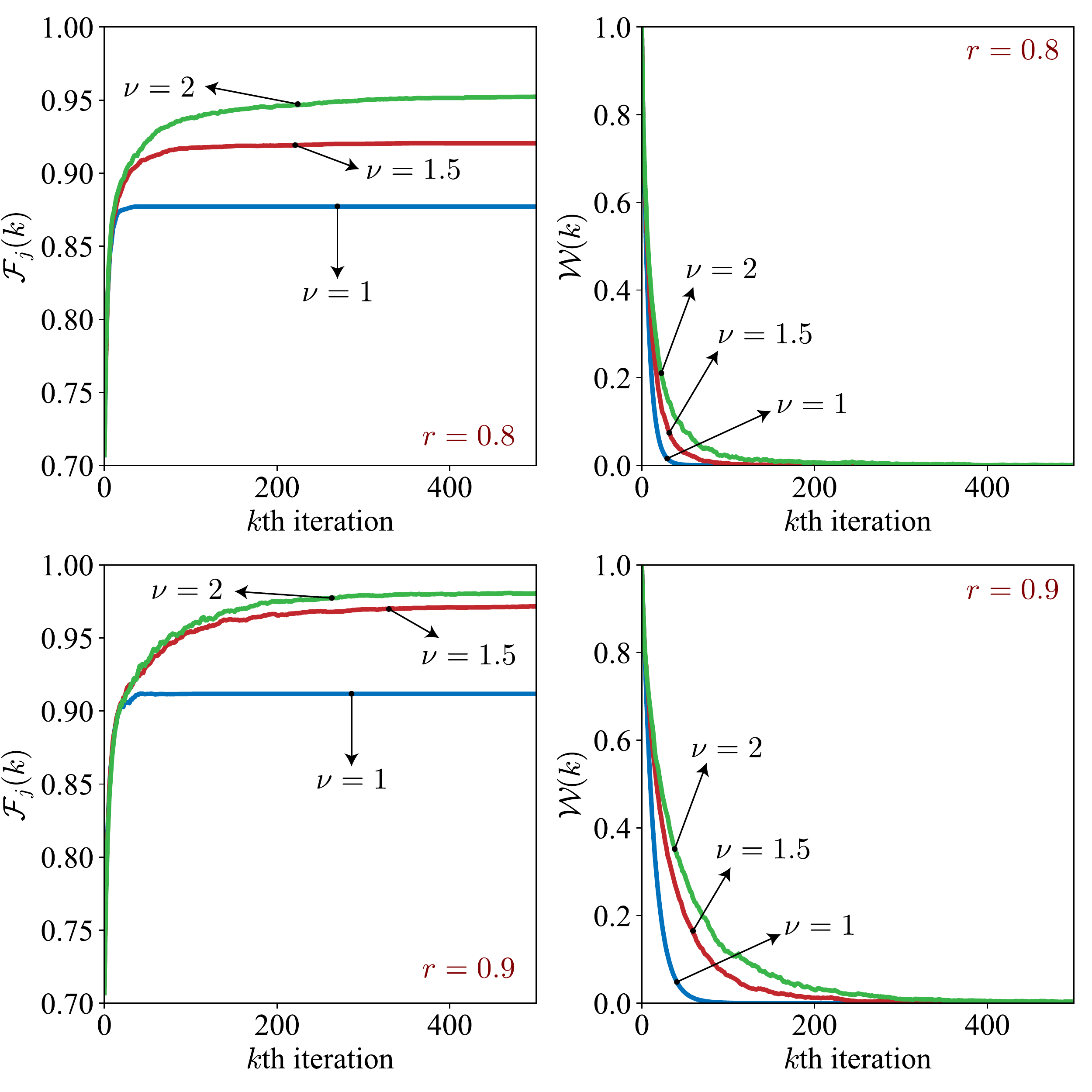}
	\caption{Numerical results for the mean fidelity $\mathcal{F}_0(k)$ and the mean searching rate $\mathcal{W}(k)$ given by Eq. (\ref{Eq33}), where $\hat{\mathcal{O}}_E=\hat{S}_x$. We employ $\mathcal{N}=1000$.}
	\label{Fig5}
\end{figure}

\begin{figure}[t]
	\centering
	\includegraphics[width=0.6\linewidth]{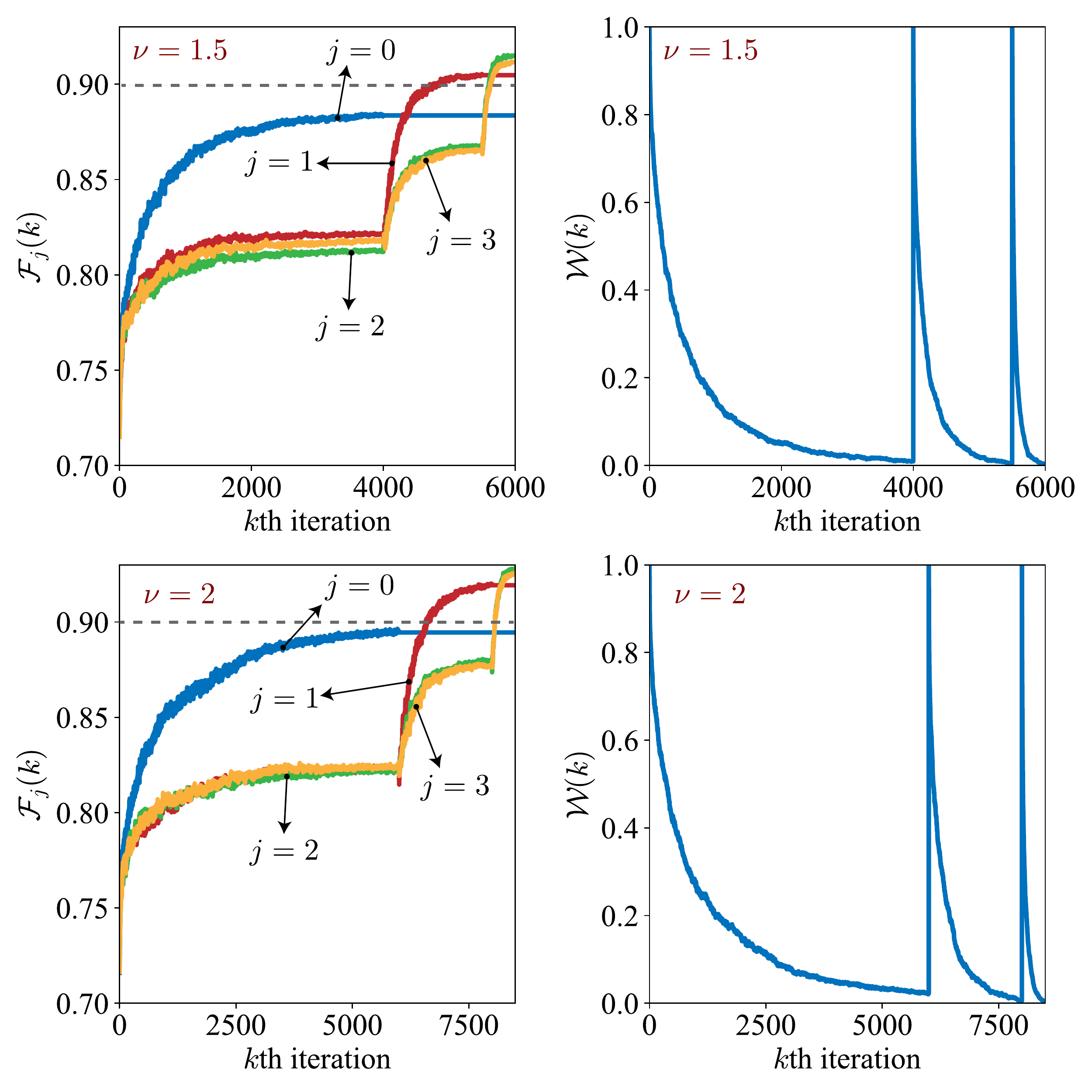}
	\caption{Numerical results for the mean fidelity $\mathcal{F}_j(k)$ and the mean searching rate $\mathcal{W}(k)$ given by Eq. (\ref{Eq33}), where $\hat{\mathcal{O}}_E$ is a random two-qubit operator. We employ $\mathcal{N}=1000$ and $r=0.9$.}
	\label{Fig6}
\end{figure}

For the general performance of our protocol, we start with a $\hat{\mathcal{O}}_E$ described by a random Hermitian matrix. Figure \ref{Fig3} shows the mean fidelity $\mathcal{F}_0(k)=\mathcal{F}_1(k)$ for different values of the reward rate $r$, and the parameter $\nu$. From this figure, we can see that for $r=0.9$ and $\nu=2$, we obtain $\mathcal{F}_0(k)>0.98$ with $k<300$. Also, in all cases we have $\mathcal{F}_0(k)>0.90$ for $k<10$. It means that using a reduced number of iterations we can obtain good fidelities for the eigenvector of a completely random single-qubit operator. On the other hand, we observe that when $r$ and $\nu$ are larger,  the maximum value of $\mathcal{F}_0(k)$ increases, but we need more iterations for the convergence of the algorithm. Figure \ref{Fig4} shows the mean searching range $\mathcal{W}(k)$ for the same cases. From this figure we can clearly see how the algorithm needs less iterations when $r$ and $\nu$ decrease, with the extreme case of $r=0.6$, $\nu=1$, where the algorithm converges before 70 iterations.

Now, we consider a particular example $\hat{\mathcal{O}}_E=\hat{S}_x=\frac{1}{2}\sigma_x$. In this case, the distance in the Bloch sphere between $\ket{0}$ and the eigenstates of $\hat{\mathcal{O}}_E$  is the largest possible. Figure \ref{Fig5} shows that our algorithm converges with few iterations to good approximations of the eigenvectors, we can see that we obtain the eigenvectors with fidelity above 98$\%$ in 400 iterations, for the case $\nu=2$ and $r=0.9$. 

As we can see, the maximum fidelity for the case $\hat{\mathcal{O}}_E=\hat{S}_x$ has decreased with respect to the random one. This is because the distance between $\ket{0}$ and the eigenvectors of $\hat{S}_x$  is larger than the distance between $\ket{0}$ and the eigenvectors of $\hat{\mathcal{O}}_E$ in the random case, therefore, the protocol has worse convergence.

\subsection{Two-qubit case}

\begin{figure}[t]
	\centering
	\includegraphics[width=0.6\linewidth]{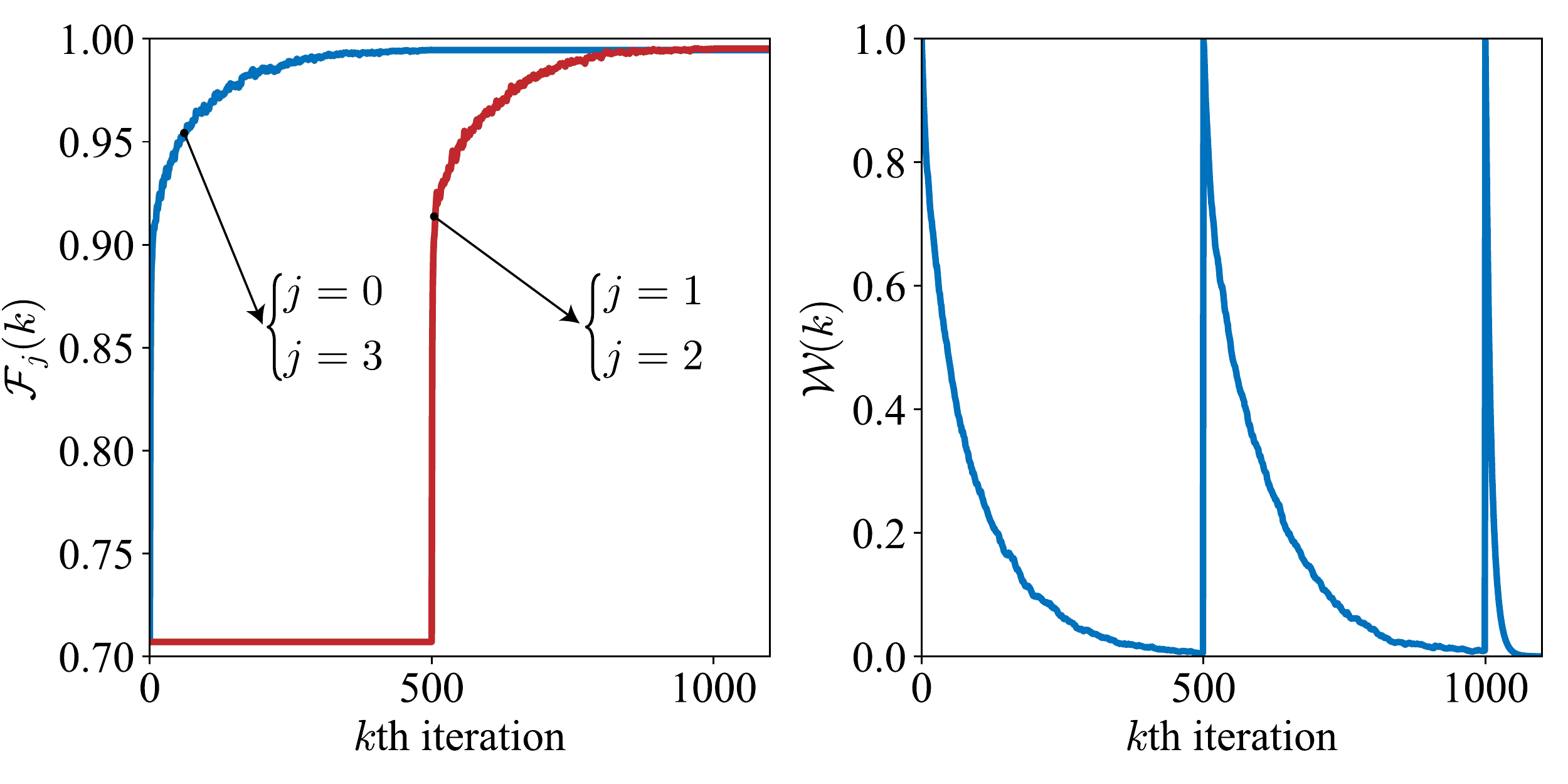}
	\caption{Numerical results for the mean fidelity $\mathcal{F}_j(k)$ and the mean searching rate $\mathcal{W}(k)$ given by Eq. (\ref{Eq33}). Here, $\hat{\mathcal{O}}_E=\hat{B}$, which is described by Eq.~(\ref{Eq34}). We employ $\mathcal{N}=1000$, $r=0.9$, and $\nu=2$.}
	\label{Fig7}
\end{figure}

This case is analogous to the single-qudit case with $d=4$. First, for a general performance, we consider $\hat{\mathcal{O}}_E$ as a random two-qubit operator. Moreover, we choose $\mathcal{N}=1000$ and calculate the mean fidelity $\mathcal{F}_j(k)$ and the mean searching range $\mathcal{W}_j$ given by Eq. (\ref{Eq33}). Figure \ref{Fig6} shows the numerical calculation for $r=0.9$ and $\nu=\{1.5,2\}$. It shows again that for small $\nu$ the convergence is faster but the maximum value of $\mathcal{F}_j$ is smaller. Furthermore, with $\nu=2$ we need $8500$ iterations such that the four approximate eigenvectors converge.  With $\nu=1.5$, we only need $6000$ iterations. Nevertheless, for $\nu=2$ we obtain $\mathcal{F}_j>0.89$ for all $j$, with even $\mathcal{F}_2$ and $ \mathcal{F}_3$ up to $0.93$. In the other case, with $\nu=1.5$, the maximum values are $\mathcal{F}_0\sim0.88$, and $\{\mathcal{F}_1,\mathcal{F}_2,\mathcal{F}_3\}<0.92$. 
Also, we can see from the evolution of $\mathcal{W}(k)$ that the number of iterations needed for the convergence is smaller each time that the algorithm starts again to approximate the next eigenvector, that is, $N_0>N_1>N_2$. Finally, we consider as special case $\hat{\mathcal{O}}_E=\hat{B}$, where $\hat{B}$ is an operator given by
\begin{equation}
	\hat{B}=\ketbra{\phi_+}{\phi_+}-\ketbra{\phi_-}{\phi_-}+2\left(\ketbra{\psi_+}{\psi_+}-\ketbra{\psi_-}{\psi_-}\right),
	\label{Eq34}
\end{equation} 
with
\begin{eqnarray}
	\ket{\phi_{\pm}}=\sqrt{\frac{1}{2}}\left(\ket{00}\pm\ket{11}\right),\nonumber\\
	\ket{\psi_{\pm}}=\sqrt{\frac{1}{2}}\left(\ket{01}\pm\ket{10}\right),
	\label{Eq35}
\end{eqnarray}
the maximally-entangled Bell states. Figure \ref{Fig7} shows the performance of our protocol for this case. We can see that we obtain high fidelities ($\mathcal{F}_j>0.99$) with only 1000 iterations to approximate the four eigenvectors. We obtain this performance due to the fact that our algorithm is sensitive to the number of the product states involved in each subspace (dimension of the subspace) and not to the total dimension of the operator $\hat{\mathcal{O}}_E$. In this case, the operator $\hat{B}$ is block-diagonal, where one block acts in the subspace $\{\ket{00},\ket{11}\}$ and the other in $\{\ket{01},\ket{10}\}$. This implies that the present case is similar to two independent single-qubit cases. In Fig. \ref{Fig7}, we can see that from $k=1$ to $k=500$  we approximate the eigenstates of the first block, that is $\ket{\phi_{\pm}}$ at the same time, and from $k=501$ to $k=1000$ we approximate the eigenstates of the second block $\ket{\psi_{\pm}}$, where both cases have a performance similar to the single-qubit case.

\section{Conclusions}
We propose and analyze an approximate quantum eigensolver based on reinforcement learning with minimal resources. This proposal can be classified as a hybrid classical-quantum algorithm, such that we use a classical optimization algorithm to change a quantum system to improve a quantum task using a feedback loop combined with partially-random unitary gates.  This is in contrast with other hybrid algorithms that measure the fidelities or some expectation value in each step. Therefore, our proposal is advantageous with respect to the usual hybrid algorithms, in the sense that our protocol needs minimal storage to save only the last step of the algorithm and employs just one single-shot measurement per iteration, instead of fidelities or expectation-value measurements, which decrease the effect of the source of noise. Moreover, our protocol considers pseudo-random two-level rotations, such that it is not necessary to implement high-fidelity operations, because the randomness of the algorithm absorbs the errors of the gates. For this reason, our algorithm would be experimentally feasible in almost any current quantum platform. 

Additionally, we validated our proposal with numerical calculations of four different choices of the operator $\hat{\mathcal{O}}_E$, random single-qubit operator, $\hat{S}_x$ operator, random two-qubit operator, and $\hat{B}$ operator defined by Eq. (\ref{Eq34}), obtaining as a general rule that our algorithm reaches higher fidelities for the approximate eigenvectors for large values of $\nu$ and $r$, but the convergence in this case is slower. This is related to the balance between exploration and exploitation typical from reinforcement learning algorithms. Moreover, our algorithm is sensitive to the size of the different subspaces expanded by product states and not to the size of the total space of the operator $\hat{\mathcal{O}}_E$. This is the case showed in Fig. (\ref{Fig7}), where the eigenvectors are the maximally-entangled Bell states. We point out that, in order to improve the performance of the protocol in future extensions, it could be interesting to study dynamical reward rates (r) and dynamical parameter $\nu$. 

Finally, due to the simplicity, minimal resources employed by our protocol, and the fact that we need only a basic classical processor (command center) capable to perform pseudo-random rotations, it can be useful for the development of near future semi-autonomous quantum devices, which will have to make decisions with incomplete information obtained by interaction with the external environment.

\ack
We acknowledge support from Financiamiento Basal para Centros Cient\'ificos y Tecnol\'ogicos de Excelencia (Grant No. FB0807),  projects QMiCS (820505) and OpenSuperQ (820363) of the EU Flagship on Quantum Technologies, EU FET Open Grant Quromorphic, Basque Government IT986-16, and PGC2018-095113-B-I00 (MCIU/AEI/FEDER, UE).

\section*{Data availability statement}
The data that support the findings of this study are openly available at https://github.com/PanchoAlbarran/EigenSolver

\newpage
\appendix

\section{Explicit form of $\bar{\theta}^{(k)}$ and $\bar{\phi}^{(k)}$} 
\label{appA}

Here, we further clarify the protocol developed in the main text.
\vspace{1cm}

From Eq. (\ref{Eq04}), we have
\begin{eqnarray}
	\ket{0}&=&\cos\left(\frac{\alpha}{2}\right)\ket{v_0}+\sin\left(\frac{\alpha}{2}\right)\ket{v_1}\nonumber\\
	\ket{1}&=&e^{-i\beta}\left[\sin\left(\frac{\alpha}{2}\right)\ket{v_0}-\cos\left(\frac{\alpha}{2}\right)\ket{v_1}\right].
	\label{EqA1}
\end{eqnarray}
Replacing Eq. (\ref{Eq07}) we obtain
\begin{eqnarray}
	\ket{\phi_{A,0}^{(k)}}&=&\left[\cos\left(\frac{\theta^{(k)}}{2}\right)\cos\left(\frac{\alpha}{2}\right)+e^{i(\varphi^{(k)}-\beta)}\sin\left(\frac{\theta^{(k)}}{2}\right)\sin\left(\frac{\alpha}{2}\right)\right]\ket{v_0}\nonumber\\
	&+&\left[\cos\left(\frac{\theta^{(k)}}{2}\right)\sin\left(\frac{\alpha}{2}\right)-e^{i(\varphi^{(k)}-\beta)}\sin\left(\frac{\theta^{(k)}}{2}\right)\cos\left(\frac{\alpha}{2}\right)\right]\ket{v_1}.\qquad
	\label{EqA2}
\end{eqnarray}
Thus,
\begin{eqnarray}
	\fl
	\ket{\bar{\phi}_{A,0}^{(k)}}&=&\hat{U}_E\ket{\phi_{A,0}^{(k)}}=e^{-i\lambda_0\tau}\left[\cos\left(\frac{\theta^{(k)}}{2}\right)\cos\left(\frac{\alpha}{2}\right)+e^{i(\varphi^{(k)}-\beta)}\sin\left(\frac{\theta^{(k)}}{2}\right)\sin\left(\frac{\alpha}{2}\right)\right]\ket{v_0}\nonumber\\
	\fl
	&+&e^{-i\lambda_1\tau}\left[\cos\left(\frac{\theta^{(k)}}{2}\right)\sin\left(\frac{\alpha}{2}\right)-e^{i(\varphi^{(k)}-\beta)}\sin\left(\frac{\theta^{(k)}}{2}\right)\cos\left(\frac{\alpha}{2}\right)\right]\ket{v_1}.
	\label{EqA3}
\end{eqnarray}
By means of the definition of $\ket{v_0}$ and $\ket{v_1}$ given by Eq. (\ref{Eq04}), we obtain
\begin{eqnarray}
	&&\ket{\bar{\phi}_{A,0}^{(k)}}=e^{-i\lambda_0\tau}\left[\cos\left(\frac{\theta^{(k)}}{2}\right)\cos\left(\frac{\alpha}{2}\right)+e^{i(\varphi^{(k)}-\beta)}\sin\left(\frac{\theta^{(k)}}{2}\right)\sin\left(\frac{\alpha}{2}\right)\right]\nonumber\\
	&&\cdot\left[\cos\left(\frac{\alpha}{2}\right)\ket{0}+e^{i\beta}\sin\left(\frac{\alpha}{2}\right)\ket{1}\right]+e^{-i\lambda_1\tau}\Bigg[\cos\left(\frac{\theta^{(k)}}{2}\right)\sin\left(\frac{\alpha}{2}\right)\nonumber\\
	&&-e^{i(\varphi^{(k)}-\beta)}\sin\left(\frac{\theta^{(k)}}{2}\right)\cos\left(\frac{\alpha}{2}\right)\Bigg]\left[\sin\left(\frac{\alpha}{2}\right)\ket{0}-e^{i\beta}\cos\left(\frac{\alpha}{2}\right)\ket{1}\right].
	\label{EqA4}
\end{eqnarray}

We rewrite the eigenvalues as $\lambda_0=\delta-\lambda$ and $\lambda_1=\delta+\lambda$ where $\delta=(\lambda_1+\lambda_0)/2$ and $\lambda=(\lambda_1-\lambda_0)/2$. Then, we rewrite Eq. (\ref{EqA4}) up to a global phase as

\begin{eqnarray}
	\fl
	&&\ket{\bar{\phi}_{A,0}^{(k)}}=\Bigg[\cos\left(\frac{\theta^{(k)}}{2}\right)\cos(\lambda\tau)+i\cos\left(\frac{\theta^{(k)}}{2}\right)\cos(\alpha)\sin(\lambda\tau)\nonumber\\
	\fl
	&&+ie^{i(\varphi^{(k)}-\beta)}\sin\left(\frac{\theta^{(k)}}{2}\right)\sin(\alpha)\sin(\lambda\tau)\Bigg]\ket{0}+e^{i\varphi^{(k)}}\Bigg[ie^{-i(\varphi^{(k)}-\beta)}\cos\left(\frac{\theta^{(k)}}{2}\right)\sin(\alpha)\sin(\lambda\tau)\nonumber\\
	\fl
	&&+\sin\left(\frac{\theta^{(k)}}{2}\right)\cos(\lambda\tau)-i\sin\left(\frac{\theta^{(k)}}{2}\right)\cos(\alpha)\sin(\lambda\tau)\Bigg]\ket{1}.
	\label{EqA5}
\end{eqnarray}
This state has the form 
\begin{equation}
	\ket{\bar{\phi}_{A,0}^{(k)}}=(a_0+ib_0)\ket{v_0}+e^{i\varphi^{(k)}}(a_1+ib_1)\ket{v_1},
	\label{EqA6}
\end{equation}
with
\begin{eqnarray}
	\fl
	a_0&=&\cos\left(\frac{\theta^{(k)}}{2}\right)\cos(\lambda\tau)-\sin(\varphi^{(k)}-\beta)\sin\left(\frac{\theta^{(k)}}{2}\right)\sin(\alpha)\sin(\lambda\tau),\nonumber\\
	\fl
	 b_0&=&\cos\left(\frac{\theta^{(k)}}{2}\right)\cos(\alpha)\sin(\lambda\tau)+\cos(\varphi^{(k)}-\beta)\sin\left(\frac{\theta^{(k)}}{2}\right)\sin(\alpha)\sin(\lambda\tau),\nonumber\\
	 \fl
	 a_1&=&\sin\left(\frac{\theta^{(k)}}{2}\right)\cos(\lambda\tau)+\sin(\varphi^{(k)}-\beta)\cos\left(\frac{\theta^{(k)}}{2}\right)\sin(\alpha)\sin(\lambda\tau),\nonumber\\
	 \fl
	 b_1&=&-\sin\left(\frac{\theta^{(k)}}{2}\right)\cos(\alpha)\sin(\lambda\tau)+\cos(\varphi^{(k)}-\beta)\cos\left(\frac{\theta^{(k)}}{2}\right)\sin(\alpha)\sin(\lambda\tau).
	 \label{EqA7}
\end{eqnarray}

Finally, up to a global phase, the state given by Eq. (\ref{EqA7}) can be written in the form of Eq. (\ref{Eq08}), where
\begin{eqnarray}
\fl
\bar{\theta}^{(k)}=\cos^{-1}\left(\sqrt{a_0^2+b_0^2}\right);\,\bar{\varphi}^{(k)}=\left[\varphi^{(k)}+\tan^{-1}\left(\frac{b_1}{a_1}\right)-\tan^{-1}\left(\frac{b_0}{a_0}\right)\right]\, \textrm{mod}(2\pi).
\end{eqnarray}

\section{Explicit form of $\Delta_{\theta}^{(k)}$ and $\Delta_{\varphi}^{(k)}$}
\label{appB}
From Eqs. (\ref{Eq07}) and (\ref{Eq09}) we have,
\begin{eqnarray}
	\ket{0}&=&\cos\left( \frac{\theta^{(k)}}{2}\right)\ket{\phi_{A,0}^{(k)}}+\sin\left( \frac{\theta^{(k)}}{2}\right)\ket{\phi_{A,1}^{(k)}},\nonumber\\
	\ket{1}&=&e^{-i\varphi^{(k)}}\sin\left( \frac{\theta^{(k)}}{2}\right)\ket{\phi_{A,0}^{(k)}}-e^{-i\varphi^{(k)}}\cos\left( \frac{\theta^{(k)}}{2}\right)\ket{\phi_{A,1}^{(k)}}.
\end{eqnarray}
Replacing this expression in the first line of Eq. (\ref{Eq08}), we obtain
\begin{eqnarray}
	\fl
	\ket{\bar{\phi}_{A,0}^{(k)}}=&&\left[\cos\left( \frac{\bar{\theta}^{(k)}}{2}\right)\cos\left( \frac{\theta^{(k)}}{2}\right)+e^{i(\bar{\varphi}^{(k)}-\varphi^{(k)})}\sin\left( \frac{\bar{\theta}^{(k)}}{2}\right)\sin\left( \frac{\theta^{(k)}}{2}\right)\right]\ket{\phi_{A,0}^{(k)}}\nonumber\\
	\fl
	&&+\left[\cos\left( \frac{\bar{\theta}^{(k)}}{2}\right)\sin\left( \frac{\theta^{(k)}}{2}\right)-e^{i(\bar{\varphi}^{(k)}-\varphi^{(k)})}\sin\left( \frac{\bar{\theta}^{(k)}}{2}\right)\cos\left( \frac{\theta^{(k)}}{2}\right)\right]\ket{\phi_{A,1}^{(k)}}\nonumber\\
	\fl
	=&&e^{i\Psi_0}\cos\left(\frac{\Delta_{\theta}^{(k)}}{2}\right)\ket{\phi_{A,0}^{(k)}}+e^{i\Psi_1}\sin\left(\frac{\Delta_{\theta}^{(k)}}{2}\right)\ket{\phi_{A,1}^{(k)}},
\end{eqnarray}
where
\begin{eqnarray}
	\fl
	\cos^2\left(\frac{\Delta_{\theta}^{(k)}}{2}\right)&=&\cos^2\left( \frac{\bar{\theta}^{(k)}}{2}\right)\cos^2\left( \frac{\theta^{(k)}}{2}\right)+\sin^2\left( \frac{\bar{\theta}^{(k)}}{2}\right)\sin^2\left( \frac{\theta^{(k)}}{2}\right)\nonumber\\
	\fl
	&+&2\cos\left( \frac{\bar{\theta}^{(k)}}{2}\right)\sin\left( \frac{\bar{\theta}^{(k)}}{2}\right)\cos\left( \frac{\theta^{(k)}}{2}\right)\sin\left( \frac{\theta^{(k)}}{2}\right)\cos(\bar{\varphi}^{(k)}-\varphi^{(k)})\nonumber\\
	\fl
	&=&\left[\cos\left( \frac{\bar{\theta}^{(k)}}{2}\right)\cos\left( \frac{\theta^{(k)}}{2}\right)+\sin\left( \frac{\bar{\theta}^{(k)}}{2}\right)\sin\left( \frac{\theta^{(k)}}{2}\right)\right]^2\nonumber\\
	\fl
	&+&2\cos\left( \frac{\bar{\theta}^{(k)}}{2}\right)\sin\left( \frac{\bar{\theta}^{(k)}}{2}\right)\cos\left( \frac{\theta^{(k)}}{2}\right)\sin\left( \frac{\theta^{(k)}}{2}\right)\left[\cos(\bar{\varphi}^{(k)}-\varphi^{(k)})-1\right]\nonumber\\
	\fl
	&=&\cos^2\left(\frac{\bar{\theta}^{(k)}-\theta^{(k)}}{2}\right)+\frac{1}{2}\sin(\bar{\theta}^{(k)})\sin(\theta^{(k)})\left[\cos(\bar{\varphi}^{(k)}-\varphi^{(k)})-1\right],
\end{eqnarray}
\begin{eqnarray}
	\fl
	\Psi_0=\tan^{-1}\left[ \frac{\sin(\bar{\varphi}^{(k)}-\varphi^{(k)})\sin\left( \frac{\bar{\theta}^{(k)}}{2}\right)\sin\left( \frac{\theta^{(k)}}{2}\right)}{\cos\left( \frac{\bar{\theta}^{(k)}}{2}\right)\cos\left( \frac{\theta^{(k)}}{2}\right)+\cos(\bar{\varphi}^{(k)}-\varphi^{(k)})\sin\left( \frac{\bar{\theta}^{(k)}}{2}\right)\sin\left( \frac{\theta^{(k)}}{2}\right)}\right]
\end{eqnarray}
and
\begin{eqnarray}
	\fl
	\Psi_1=\tan^{-1}\left[ \frac{\sin(\bar{\varphi}^{(k)}-\varphi^{(k)})\sin\left( \frac{\bar{\theta}^{(k)}}{2}\right)\cos\left( \frac{\theta^{(k)}}{2}\right)}{\cos\left( \frac{\bar{\theta}^{(k)}}{2}\right)\sin\left( \frac{\theta^{(k)}}{2}\right)+\cos(\bar{\varphi}^{(k)}-\varphi^{(k)})\sin\left( \frac{\bar{\theta}^{(k)}}{2}\right)\cos\left( \frac{\theta^{(k)}}{2}\right)}\right].
\end{eqnarray}
Finally, up to a global phase, we can write the state $\ket{\bar{\phi}^{(k)}_{A,0}}$ as 
\begin{equation}
	\ket{\bar{\phi}_{A,0}^{(k)}}=\cos\left(\frac{\Delta_{\theta}^{(k)}}{2}\right)\ket{\phi_{A,0}^{(k)}}+e^{i\Delta_{\phi}^{(k)}}\sin\left(\frac{\Delta_{\theta}^{(k)}}{2}\right)\ket{\phi_{A,1}^{(k)}}
\end{equation}
with $\Delta_{\phi}^{(k)}=\Psi_1-\Psi_0$.


\end{document}